# A Pneumatic Chaotic Pendulum

Devlin Gualtieri
Tikalon LLC, Ledgewood, New Jersey
(gualtieri@ieee.org)

*I present a chaotic pendulum based on the repulsive force between a random array of point sources of air flow and the conical tip of a rigid pendulum. Source code is provided for generation of random aperture arrays. The chaotic motion was analyzed using machine vision techniques. A computer simulation of this system is also presented, as are the results of some example simulations.*

Introduction

There are numerous examples of chaotic pendulums based on repulsive forces between an array of permanent magnets and a rigid rod tipped with a permanent magnet.[1-2] Air flow will also provide a repulsive lateral force on a pendulum when acting on a cone. A pneumatic chaotic pendulum was constructed using an air plenum drilled with an array of randomly-sized holes in random positions. A diagram of its construction can be viewed as fig. 1.

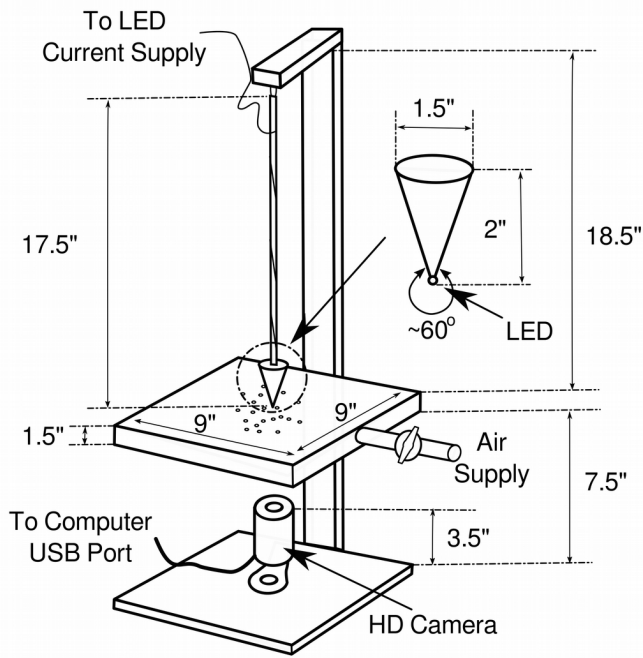

Fig. 1. The pneumatic chaotic pendulum.

The 9-inch by 9-inch air plenum is a frame that's capped at the top and bottom by plexiglas polycarbonate sheets. The sheet transparency allows the tip of the pendulum to be viewed from below. The aperture array is constrained to a 4-inch by 4-inch patch in the center of the top sheet.

Fine wires are wrapped around the ceramic pendulum rod to supply electrical current to the LED at the cone tip.

Air Plenum

The air plenum is constructed as a 9-inch by 9-inch frame that's capped at the top and bottom by transparent plexiglas polycarbonate sheets. The aperture array of pnematic point sources is constrained to a 4-inch by 4-inch patch in the center of the top sheet. Holes were drilled through this sheet in a range of sizes from 0.0980-inch to 0.1495-inch (ANSI drill sizes 40 to 25). A computer program (source code in Appendix II) produced a drilling template for twenty holes of random size in random positions, as shown in fig. 2.

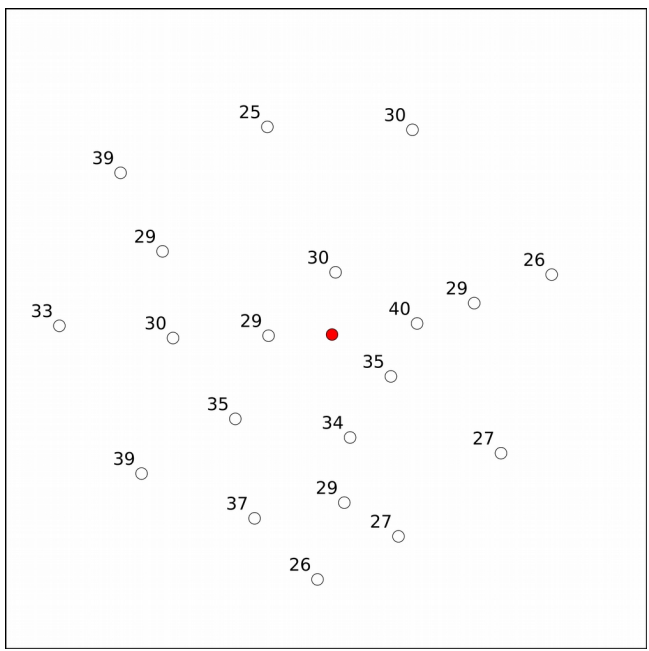

Fig. 2. Drilling template for the top sheet of the air plenum. The numbers are the ANSI drill sizes, and the red dot marks the center of the array.

Pendulum

Since air flow was limited (a Porter Cable Model C2002 Portable Electric Pancake Air Compressor with an airflow of about 3 cubic feet per minute was used), it was necessary to fabricate a pendulum of small weight. The cone was hollow, constructed from paper, and it was mounted at the end of a thin, 0.125 inch diameter, ceramic rod manufactured as a feed-through for a fine gauge thermocouple. Alternatively, a hollow but rigid plastic tube can be used. This might be constructed from several plastic drinking straws coupled together.

A number of cone angles was tried, and maximum motion was obtained with a cone of about 60-degree tip angle. A small light-emitting diode was affixed to the tip of the cone to assist in the machine vision data collection, as described below. The combined weight of the rod, cone, LED, and LED wiring was 9.0 grams, and the rigid rod was attached to the support with a short cotton thread.  It was important to adjust the rate of airflow with the valve, since too high a flow resulted in the cone's orbiting the entire array.

Machine Vision Data Collection

To facilitate data collection, the motion of the tip of the pendulum was recorded on video for analysis.  This process was aided by fitting the tip of the cone with a small light-emitting diode operating at low power and collecting data in a darkened room.  A small HD resolution video camera with a USB interface was used with the free and open source video capture program, *cheese*.  Although the camera was capable of color video at 1920 x1080 resolution, the data collection was done using grayscale at 640 x 480.

The video stream, captured as a *webm* video file, was processed by the free and open source video editing program, *ffmpeg*, to create sequential grayscale *pgm* images at half-second intervals.  An image analysis program extracted the (x,y) coordinates of the tip as a function of time.  Fig. 3 shows the (x,y) position as pixels from the pendulum rest position at half-second intervals for twelve minutes of operation after an initial 30-second period to eliminate any transient effects.  Note the unexplained region at larger x and y values that might relate to the placement and sizes of the apertures.

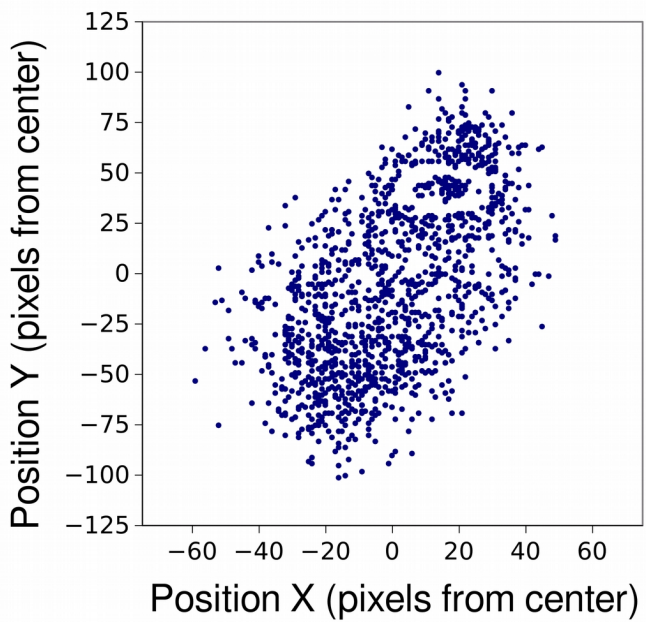

Fig. 3.  The pendulum tip position, as pixels from the pendulum rest position, at half-second intervals for twelve minutes of operation. Data for the initial 30-seconds were eliminated to exclude potential transient effects.

Note the unexplained region at larger x and y values that might relate to the placement and sizes of the apertures, as would the shape of the region of allowed values.

Data Anaysis

The radial deviation from the pendulum rest position was calculated from the (x,y) coordinates, and a middle portion of the movement is shown as fig. 4.

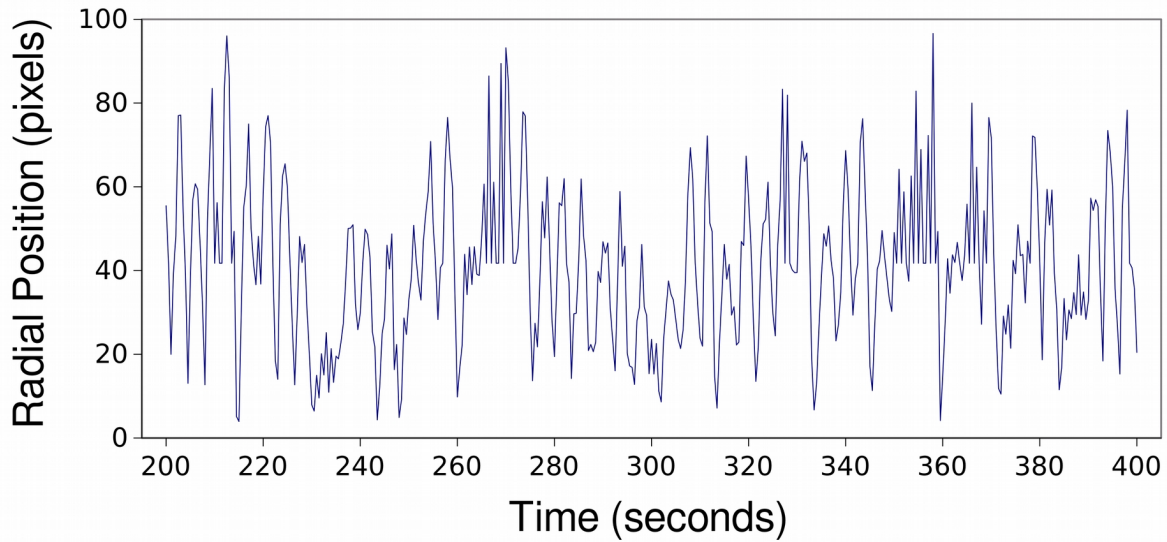

Fig. 4. Radial deviation from the pendulum rest position as a function of time. Shown are data from 200 to 400 seconds.

Summary of Experiment

This proof-of-principle experiment opens an opportunity to explore the affect of the aperture number, sizes, and positions on the pendulum motion. The mechanics of the system are simple, so a computer simulation was undertaken (Source code in Appendix III). This simulation reproduces the action of this mechanical system, and it allowed a more thorough analysis of its performance. The simulation illustrates a sensitivity to initial conditions, in this case the starting position of the pendulum tip, that's common in chaotic systems.

Physical Model

A diagram of the forces acting upon the pendulum's conical tip is shown as fig. 5. The pendulum is always acted upon by a restoring force to its rest position. Its conical tip can also be acted upon by multiple air flow sources depending on its position relative to the apertures drilled into the air plenum at random positions.

The acceleration and direction of the pendulum are calculated from the sum of the resolved force components of these forces in the x- and y-directions, and the pendulum position is calculated from these accelerations after each time tick, δt. Some simplification allowed an easier computation. Air flow through the plenum apertures was considered to be laminar, and the resultant force was considered to be acting at a point. The resolved force components depend on the angle of air flow with respect to the cone. This angle is principally the angle of the cone, but it will change with the pendulum position. In these calculations, just the cone angle was used.

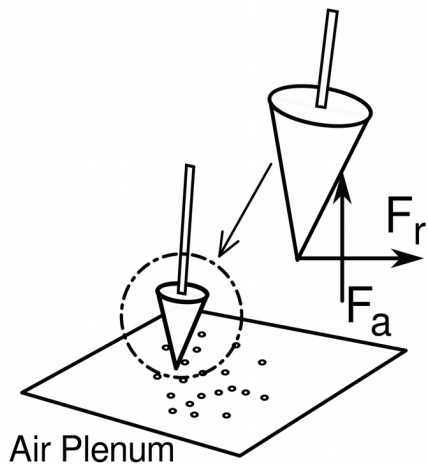

Air Plenum

Fig. 5. Forces acting on the conical tip of the pneumatic chaotic pendulum.

The air plenum is drilled with random-sized holes in random positions. The pendulum is acted upon by a restoring force to its rest position **Fr** and also forces from air flow **Fa** from the plenum apertures.

Pendulum acceleration and direction are calculated from the resolved force components in the x- and y-directions, and the pendulum position is calculated from these accelerations after each time tick, δt.

Simulation Program

A synopsis of the simulation program follows:
1) An array of randomly-sized holes at random position is generated in the air plenum.
    - A minimum spacing is required to exist between the holes.
2) A random initial pendulum position (x,y) is selected within the aperture array.
3) The force vectors acting on the conical tip are calculated:
    - The pendulum restoring force to its resting position.
    - The air forces from all apertures within the cone's projected cross-section. For simplification, a circular cross-section is used instead of the slightly elliptical one that would be present.
4) The x- and y- components of the forces are summed.
5) The resulting acceleration is calculated and summed with the last acceleration.
6) A time tick δt is set.
7) The resultant displacement vector is calculated from $\delta s = \frac{1}{2} a \, \delta t^2$.

8) A new pendulum position (x,y) is calculated
9) Loop to step 3.

The program logs the initial conditions, and the pendulum position at each clock tick. There is also provision for automatic generation of an (x,y) plot using GNUPlot (Appendix IV).[3]

Simulation Results

Fig. 6 illustrates the results of the simulation for various aperture arrays and pendulum initial positions. In some cases, the pendulum's initial position causes it to swing freely without any influence from the aperture array.

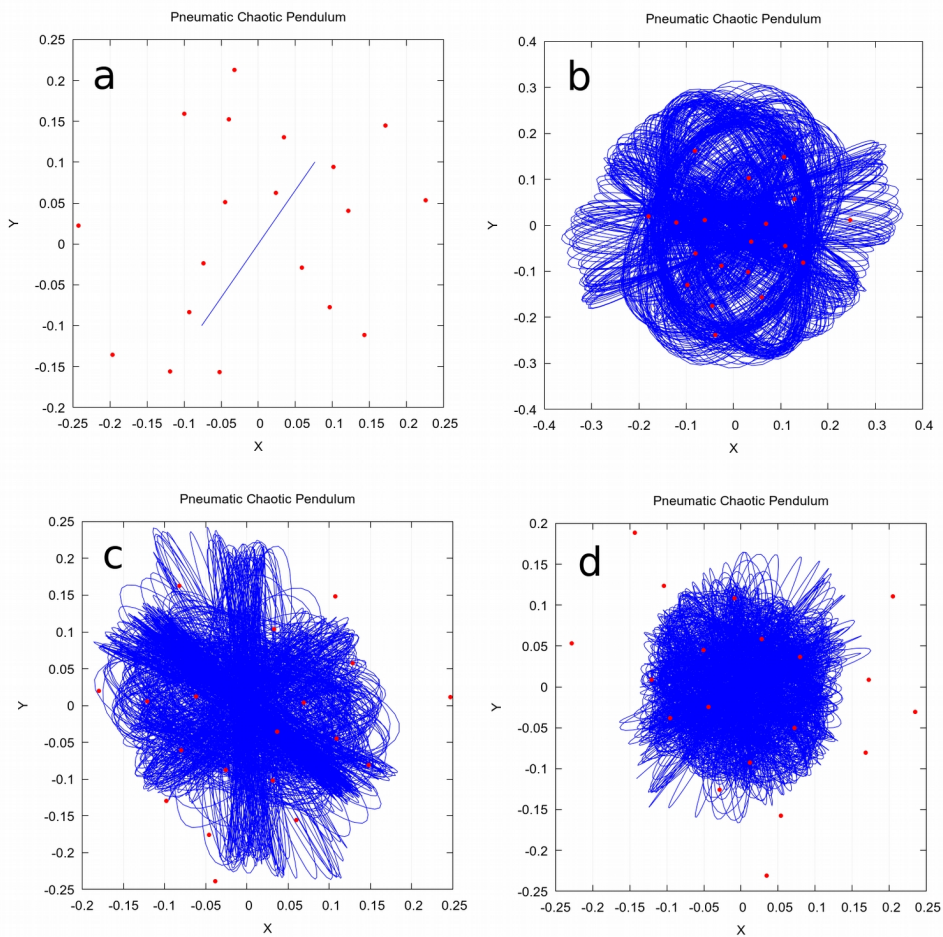

Fig. 6. Examples of the pendulum motion for different aperture arrays (shown in red) and initial pendulum positions. In (a), the pendulum swings without influence of any of the air sources, while (b) through (d) are the usual response.

Fig. 7 illustrates the temporal progression of the pendulum cone radial distance from the pendulum resting position, as sampled at one second intervals over a 1000 second period, for one set of initial conditions. Fig. 8 shows a histogram of the difference in the radial positions at 0.05 second intervals for the same initial conditions that produced the pendulum motion shown in fig. 3. The line in fig. 8 is a fit of the data to a normal distribution.

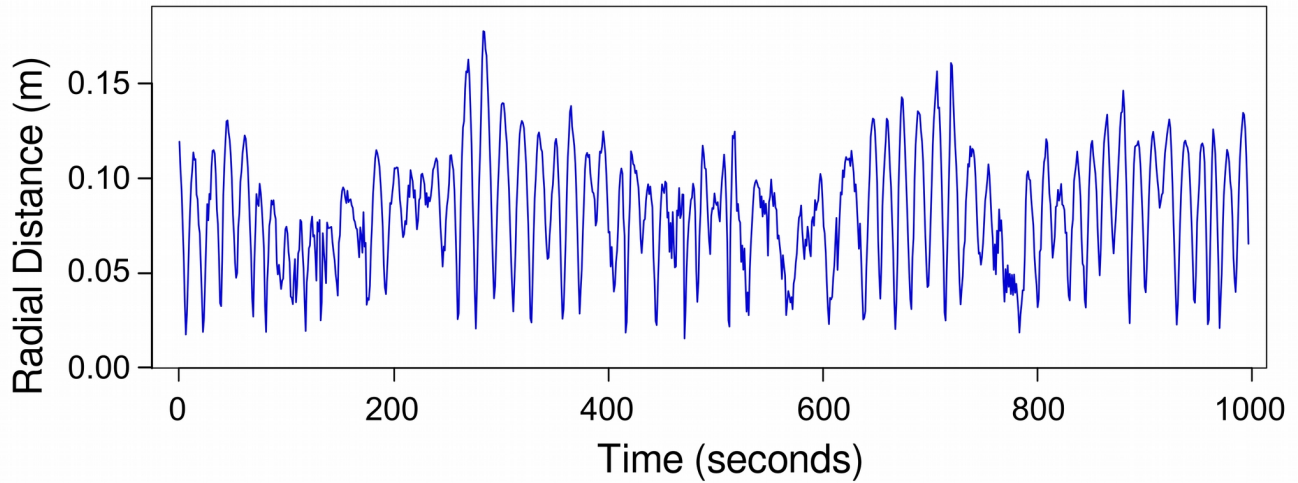

Fig. 7. Radial distance of the pendulum's conical tip from the pendulum resting position, sampled at one second intervals.

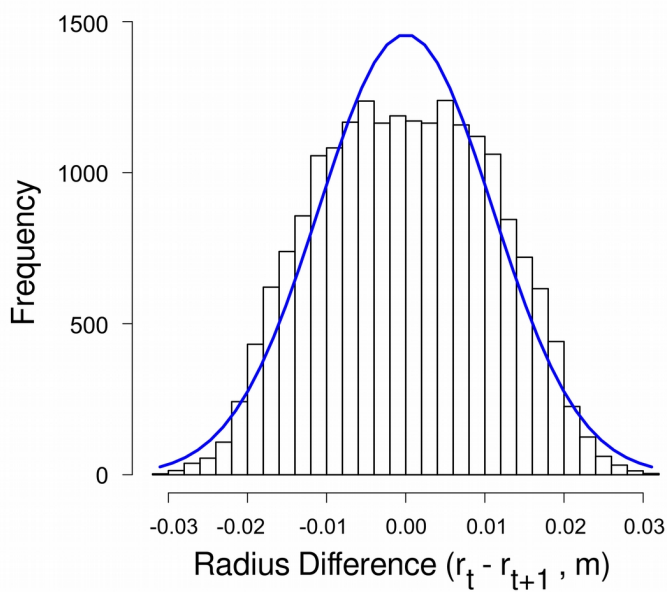

Fig. 8. Histogram of the difference in radial positions at 0.05 second intervals from the pendulum resting point for the pneumatic chaotic pendulum.

The line is a fit of the data to a normal distribution, and it indicates a diminished probability of circular orbital arcs around the pendulum resting position.

Fig. 9 shows the histograms of the pendulum x- and y- values sampled at 50 second intervals for a representative simulation of 100,000 seconds duration. The lines are fits to normal distributions.

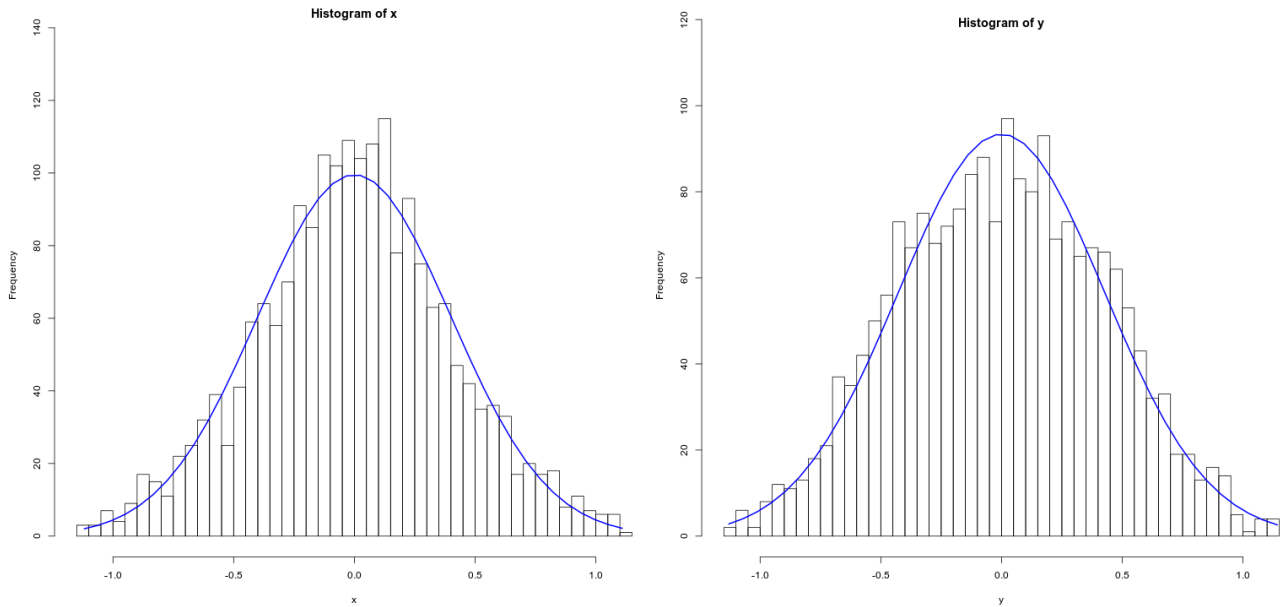

Fig. 9. Left, histogram of x-positions, and, right, histogram of y-positions, taken at 50 second intervals over the course of a representative 100,000 second simulation. The lines are fits to normal distributions.

Sensitivity to Initial Pendulum Position

Sensitivity of initial condition, the so-called *butterfly effect*, is a common feature of chaotic systems, and it's illustrated by this simulation program, which is designed to replay modified initial states. Fig. 10 is the temporal progression of the pendulum cone radial distance from the pendulum resting position, as sampled at 0.05 second intervals over a period of one minute, for two slightly different initial positions of the conical tip. In one case, the position was (-0.0800000, 0.0900000), and in the other case, the position was (-0.0799999, 0.0899999).

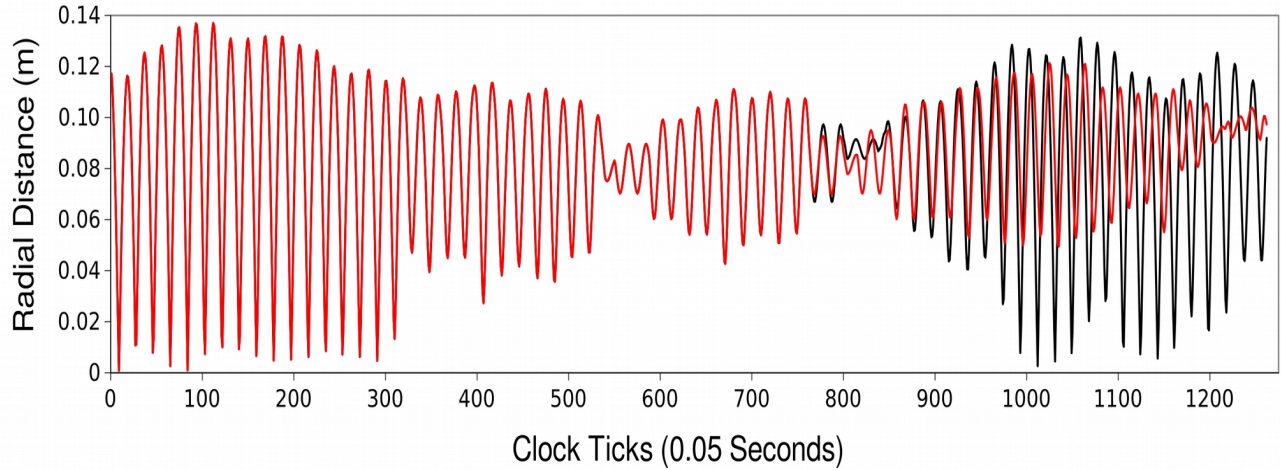

Fig. 10. Radial distance of the pendulum's conical tip from the pendulum resting position, sampled at 0.05 second intervals, for two slightly different initial positions of the pendulum tip. The cause of the divergence at about 38 seconds is explained in the text.

The divergence in this case happens as the pendulum comes under the influence of the air flow through a plenum aperture at (-0.0799045, -0.0609426). As listed in Table I, one case of the initial position leads to four interactions of the air stream with the cone, and the other results in three.

The pneumatic pendulum moves in various directions in the (x,y)-plane, but its period is fairly constant. This is confirmed in fig. 11, which plots the discrete Fourier transform of the radial motion at various times.

Table I. Interactions between the pendulum cone and air flow through a plenum aperture at (-0.0799045, -0.0609426). In one case of initial pendulum position, there are four interactions, and in the other just three. This causes the divergence at 38 seconds as shown in fig. 10.

| Initial (x,y) = (-0.0799999, 0.0899999) | | Initial (x,y) = (-0.0800000, 0.0900000) | |
| --- | --- | --- | --- |
| Cone-x | Cone-y | Cone-x | Cone-y |
| -0.086756 | -0.048205 | -0.086783 | -0.048111 |
| -0.075026 | -0.054829 | -0.075056 | -0.05473 |
| -0.062301 | -0.061284 | -0.062318 | -0.061194 |
| -0.020416 | -0.069956 | | |

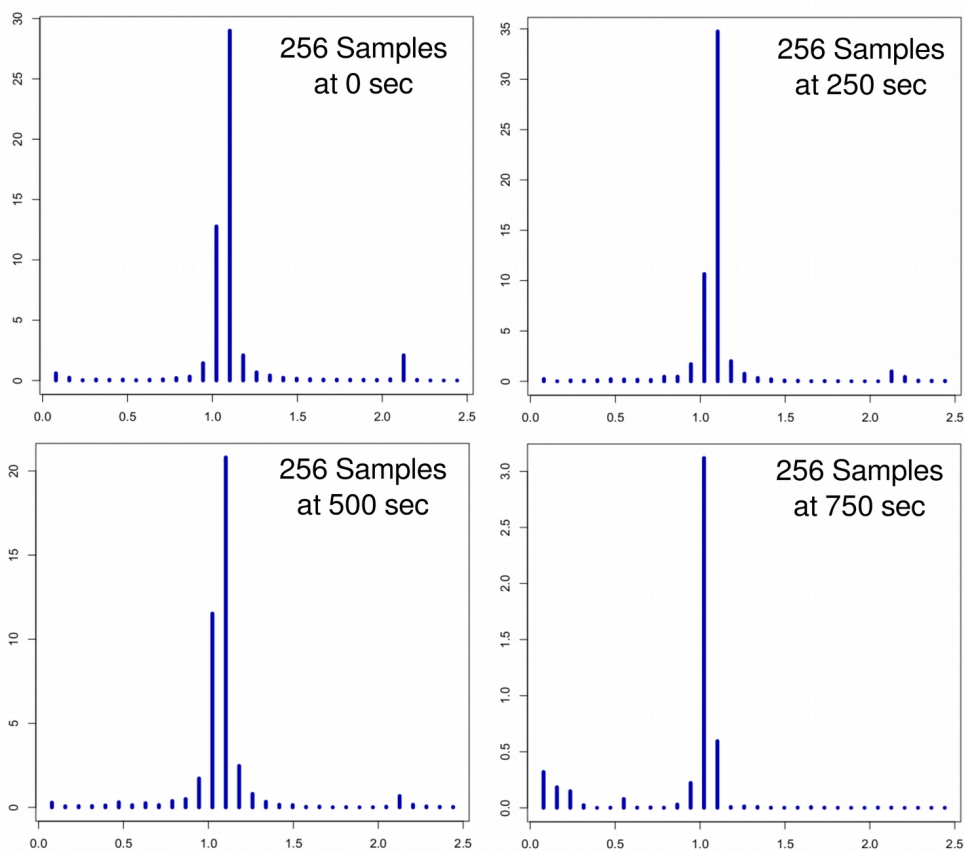

Fig. 11. Discrete Fourier transform (DFT), DC component removed, of the temporal progression of the pendulum cone radial distance from the pendulum resting position at several portions of a representative simulation (see Appendix V for the simulation parameters). This demonstrates that the pendulum period is about a second with some variation.

Supplementary Materials for Computer Simulation

Materials supplementary to this paper include the simulation program source code, and a script file for plotting with GNUPlot, both included as appendices. An MP4 video file demonstrating pendulum motion is available from the author.

3. GNUPlot, described at http://www.gnuplot.info.

Appendix I – Photographs of the apparatus for the pneumatic chaotic pendulum proof-of-concept experiment and the pendulum cone.

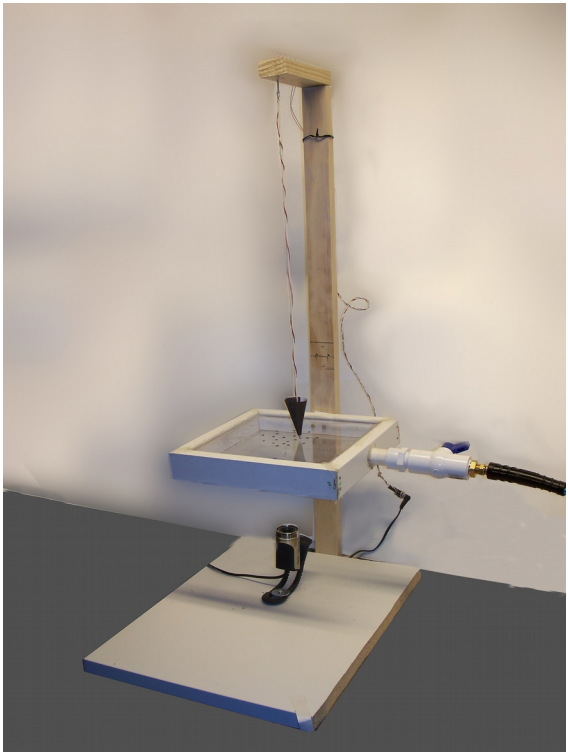 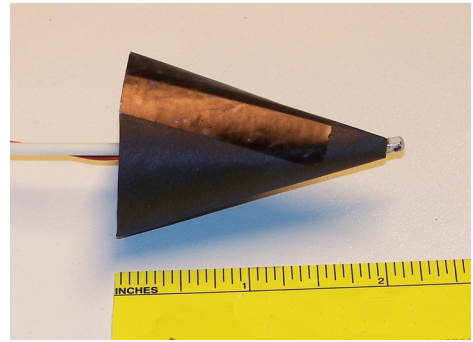

Appendix I. Left, the pneumatic chaotic pendulum; and, right, the pendulum cone.

Appendix II – Source code, pneumatic_chaotic_pendulum.c. This program generates an SVG image that can be used as a drilling template.

```c
/* -*- Mode: C; indent-tabs-mode: t; c-basic-offset: 4; tab-width: 4 -*-  */
/*
 * pneumatic_chaotic_pendulum.c
 * Copyright (C) 2020 TikalonLLC <gualtieri@ieee.org>
 *
 * pneumatic_chaotic_pendulum is free software: you can redistribute it and/or modify it
 * under the terms of the GNU General Public License as published by the
 * Free Software Foundation, either version 3 of the License, or
 * (at your option) any later version.
 *
 * pneumatic_chaotic_pendulum is distributed in the hope that it will be useful, but
 * WITHOUT ANY WARRANTY; without even the implied warranty of
 * MERCHANTABILITY or FITNESS FOR A PARTICULAR PURPOSE.
 * See the GNU General Public License for more details.
 *
 * You should have received a copy of the GNU General Public License along
 * with this program.  If not, see <http://www.gnu.org/licenses/>.
 */

#include <stdio.h>
#include <stdlib.h>
#include<time.h>
#include <math.h>

#define num_apertures 20
#define trials 1000

/* Prototypes */
char *strcpy (char *dest, const char *src);
char *strcat (char *dest, const char *src);
void exit (int status);
void srand (unsigned int seed);
int rand ();
float random_angle ();
float random_radius ();
int random_hole ();
int check_spacing (float min_spacing, float x_trial, float y_trial, int ap,
                    float x[], float y[]);
/* end of prototypes */

char fn1[64];
char fn2[64];
char st_format[255] = "";
FILE *outdata;
int ap = 0;
int i;
float x_trial, y_trial;
float x[num_apertures];
float y[num_apertures];
int drill_size[num_apertures];
float theta;
float radius;
float min_spacing = 0.2;
int xx;
int yy;
int r;

/*
Drill sizes, inch diameter
40      0.0980
39      0.0995
38      0.1015
37      0.1040
36      0.1065
35      0.1100
```

```
   34     0.1110
   33     0.1130
   32     0.1160
   31     0.1200
   30     0.1285
   29     0.1360
   28     0.1405
   27     0.1440
   26     0.1470
   25     0.1495
*/

float
random_angle ()
{
//returns a random angle in radians from 0 to 360 degrees
   return (2 * 3.1415926535) * (rand () / (float) RAND_MAX);
}

float
random_radius ()
{
//returns a random radius from 0 to 1
   return (rand () / (float) RAND_MAX);
}

int
random_hole ()
{
//returns a random drill size from 25 to 40
   return (25 + (rand () % 16));
}

int
check_spacing (float min_spacing, float x_trial, float y_trial, int ap,
               float x[], float y[])
{
//flags spacings less than minimum.  A return value of zero indicates an OK spacing
   int j;
   float sp;

   if (ap == 0)
     {
       return 0;
     }

   for (j = 0; j <= ap; j++)
     {
       sp =
         ((x[j] - x_trial) * (x[j] - x_trial)) +
         ((y[j] - y_trial) * (y[j] - y_trial));

       if (sp < (min_spacing * min_spacing))
         {
           return 1;
         }

     }

   return 0;
}

int
main (int argc, char *argv[])
{

   strcpy (fn1, "output.txt");
   strcpy (fn2, "output.svg");

   printf ("\nOutput files selected = %s\t%s\n", fn1, fn2);
```

```c
   if ((outdata = fopen (fn1, "w")) == NULL)
      {
        printf ("\nData output file cannot be opened.\n");
        exit (1);
      }

   srand (time (NULL));

   ap = 0;
   i = 0;

   while (ap < num_apertures)
      {
        theta = random_angle ();
        radius = random_radius ();
//radius squared to ensure uniform density across area
        x_trial = radius * radius * cos (theta);
        y_trial = radius * radius * sin (theta);

//check_spacing returns zero if spacing is OK
        if (check_spacing (min_spacing, x_trial, y_trial, ap, x, y) == 0)
         {
           x[ap] = x_trial;
           y[ap] = y_trial;
           drill_size[ap] = random_hole ();
           fprintf (outdata, "%4f3\t%4f3\t%d\n", x[ap], y[ap], drill_size[ap]);
           ap++;
           i++;
           if (i > trials)
              {
                printf ("\n\n%s",
                      "***Minimum spacing appears to be too large ");
                printf ("%s\n\n", "for number of apertures***");
                break;
              }
         }

      }
   fclose (outdata);

//Open SVG file

   if ((outdata = fopen (fn2, "w")) == NULL)
      {
        printf ("\nSVG output file cannot be opened.\n");
        exit (1);
      }
// Print svg header
   printf (" <svg xmlns=\"http://www.w3.org/2000/svg\" version=\"1.1\">\n");
   fprintf (outdata,
          " <svg xmlns=\"http://www.w3.org/2000/svg\" version=\"1.1\">\n");

/*
Basic SVG circle code:
<circle cx="600" cy="200" r="100" fill="red" stroke="blue" stroke-width="10"  />

Basic SVG text:
  <g font-size="30" font="sans-serif" fill="black" stroke="none"
   text-anchor="middle">
    <text x="200" y="400" dx="-30">Test</text>
  </g>
*/

//Print background

   strcat (st_format, "<rect x=\"%d\" y=\"%d\" width=\"%d\" height=\"%d\"");
   strcat (st_format,
         " fill=\"white\" stroke=\"black\" stroke-width=\"2\" />\n");
   printf ("%s", st_format);
   fprintf (outdata, st_format, 0, 0, 1100, 1100);
```

```c
//mark center
//set circle center
  xx = 550;
  yy = 550;
  r = 10;

  strcpy (st_format, "");
  strcat (st_format, "<circle cx=\"%d\"  cy=\"%d\" r=\"%d\" fill=\"red\"");
  strcat (st_format, " stroke=\"black\" stroke-width=\"1\" />\n");
  printf (st_format, xx + r, yy + r, r);
  fprintf (outdata, st_format, xx + r, yy + r, r);
  strcpy (st_format, "");

  for (i = 0; i < ap; i++)
    {
      xx = 500 * (1 + x[i]);
      yy = 500 * (1 - y[i]);
      r = 10;

      strcpy (st_format, "");
      strcat (st_format,
              "<circle cx=\"%d\"  cy=\"%d\" r=\"%d\" fill=\"white\"");
      strcat (st_format, " stroke=\"black\" stroke-width=\"1\" />\n");
      printf (st_format, xx + 50 + r, yy + 50 + r, r);
      fprintf (outdata, st_format, xx + 50 + r, yy + 50 + r, r);
      strcpy (st_format, "");
      strcat (st_format,
              "<g font-size=\"30\" font=\"sans-serif\" fill=\"black\"");
      strcat (st_format, " stroke=\"none\"  text-anchor=\"middle\">");
      strcat (st_format,
              " <text x=\"%d\" y=\"%d\" dx=\"-30\">%d</text> </g>\n");
      printf (st_format, xx + 50 - (r / 2), yy + 50 - (r / 2), drill_size[i]);
      fprintf (outdata, st_format, xx + 50 - (r / 2), yy + 50 - (r / 2),
               drill_size[i]);
    }

// Print svg footer
  printf ("</svg>");
  fprintf (outdata, "</svg>");

  fclose (outdata);

  printf ("Done\n");
return (0);
}
```

# Appendix III – Program Source Code (chaotic_pendulum_simulation.c)

```c
/* -*- Mode: C; indent-tabs-mode: t; c-basic-offset: 4; tab-width: 4 -*-  */
/*
 * pneumatic_chaotic_pendulum.c
 * Copyright (C) 2020 TikalonLLC <gualtieri@ieee.org>
 *
 * pneumatic_chaotic_pendulum is free software: you can redistribute it and/or modify it
 * under the terms of the GNU General Public License as published by the
 * Free Software Foundation, either version 3 of the License, or
 * (at your option) any later version.
 *
 * pneumatic_chaotic_pendulum is distributed in the hope that it will be useful, but
 * WITHOUT ANY WARRANTY; without even the implied warranty of
 * MERCHANTABILITY or FITNESS FOR A PARTICULAR PURPOSE.
 * See the GNU General Public License for more details.
 *
 * You should have received a copy of the GNU General Public License along
 * with this program.  If not, see <http://www.gnu.org/licenses/>.
 */

 /*
    Program synopsis:
    1) Generate random aperture array
    2) Set initial pendulum position (x,y)
    3) Calculate pendulum restoring force vector
    4) Determine aperture sources within cone cross-section
    5) Calculate pneumatic force vectors acting on pendulum from all apertures
    6) Sum force vectors
    7) Sum current acceleration with last acceleration
    8) Tick delta-T
    9) Get displacement vector from s = (1/2)a t^2
    10) Get new (x,y)
    11) Loop to (3)
  */

#include <stdio.h>
#include <stdlib.h>
#include<time.h>
#include <math.h>

#define FALSE 0
#define TRUE 1
#define validation FALSE   //set true to validate pendulum motion for no airflow
#define plot TRUE          //set to true for automatic plots by GNUPlot
#define iterations 10
#define trials 20000

#define apertures 20
#define attempts 1000
#define pendulum_length 0.445    //meters
#define aperture_array_radius 0.25      //meters
#define pendulum_mass 0.009       //kilograms
#define r_cone 0.019              //meters
#define cone_half_angle 30//degrees
#define g 9.8               //meters/sec/sec
#define plenum_pressure 102500   //pa
#define atomos_pressure 101000   //pa
#define air_density 1.225 //kg/m^3
#define tick 0.05          //time tick in seconds

/* Prototypes */
char *strcpy (char *dest, const char *src);
char *strcat (char *dest, const char *src);
int strcmp (const char *str1, const char *str2);
void exit (int status);
void srand (unsigned int seed);
int rand ();
double random_angle ();
```

```c
double random_radius ();
double random_hole ();
int check_spacing (double min_spacing, double x_trial, double y_trial,
                   int ap);
double get_distance (double x1, double y1, double x2, double y2);
size_t strftime (char *s, size_t max, const char *format,
                 const struct tm *tm);
/* end of prototypes */

char fn1[64] = "output.txt";
char fn2[64] = "";
char st_format[255] = "";
char str[128] = "";
char sys_cmd[64] = "";
FILE *outdata;
FILE *indata;
int ap = 0;
int i, j, k;
int retval;
int num_apertures;
double initial_x = 0.1;            //meters
double initial_y = 0.1;            //meters
double x_trial, y_trial;
double theta;
double radius;
double min_spacing = 0.2;
double x2;
double y2;
double x_dat;
double y_dat;
double h_dat;
int r;
int fctr;

time_t t;
struct tm *tmp;
char file_timestamp[64];
char last_file_timestamp[64] = "";

//struct definition of aperture locations and hole sizes
struct a_pos_size
{
  double x, y, hole_size;
};

struct a_pos_size aperture[64];  //global declaration of aperture array

// struct definition for force vectors
struct vect
{
  double fx, fy, dx, dy;
};

struct vect force_vector[5];     //global declaration of force vectors

// struct definition for next (x,y)
struct x_y_next
{
  double x, y;
};

struct x_y_next next[1];   //global declaration of next coordinate

// struct definition for last acceleration
struct a_last
{
  double x, y;
};

struct a_last last_a[1];   //global declaration of next coordinate
```

```c
double
random_angle ()
{
//returns a random angle in radians from 0 to 360 degrees
  return (2 * 3.1415926535) * (rand () / (double) RAND_MAX);
}

double
random_radius ()
{
//returns a random radius from 0 to 1
  return (rand () / (double) RAND_MAX);
}

double
random_hole ()
{
//calculates a random hole size from 0.100 to 0.150 inch
//returns hole size in meters
  return 0.0254 * (0.00001 * (10000 + (rand () % 5001)));
}

int
check_spacing (double min_spacing, double x_trial, double y_trial, int ap)
{
//flags spacings less than minimum.  A return value of zero indicates an OK spacing
  int j;
  double sp;

  if (ap == 0)
    {
      return 0;
    }

  for (j = 0; j <= ap; j++)
    {
      sp =
        ((aperture[j].x - x_trial) * (aperture[j].x - x_trial)) +
        ((aperture[j].y - y_trial) * (aperture[j].y - y_trial));

      if (sp < (min_spacing * min_spacing))
        {
          return 1;
        }

    }

  return 0;
}

double
get_distance (double x1, double y1, double x2, double y2)
{
//return distance between two points
  return sqrt (((x2 - x1) * (x2 - x1)) + ((y2 - y1) * (y2 - y1)));
}

double
restoring_force (void)
{
  double offset;
//F=-mgsin(theta)
//sin is approximately pendulum offset divided by pendulum length
  offset = sqrt (((next[0].x) * (next[0].x)) + ((next[0].y) * (next[0].y)));

  return pendulum_mass * g * (offset / pendulum_length);
}

double
get_air_force (double h_size)
{
//double exit_velocity = sqrt(2*(plenum_pressure-atomos_pressure)/air_density);
```

```c
   double area = 3.14159 * h_size * h_size / 4;
//return air_density*area*exit_velocity*exit_velocity*sin(3.14159*cone_half_angle/180);
   return 2 * area * (plenum_pressure -
                   atomos_pressure) * sin (3.14159 * cone_half_angle /
                                           180);;
}

void
get_forces (void)
{
//stores pendulum restoring force based on position
//checks which apertures are exerting force on cone_half_angle
  int i;
  double h;                //hypotneuse of dx-dy triangle for calculating cos and sin
//first, clear old data
  for (i = 0; i < 5; i++)
    {
      force_vector[i].fx = 0;
      force_vector[i].fy = 0;
      force_vector[i].dx = 0;
      force_vector[i].dy = 0;
    }
//index [0] is pendulum restoring force
//make sure accelerations are in the right direction
  force_vector[0].dx = -next[0].x;
  force_vector[0].dy = -next[0].y;
  h = get_distance (next[0].x, next[0].y, 0, 0);
//cos is dx/h. sin is dy/h
  force_vector[0].fx = restoring_force () * (force_vector[0].dx / h);    //scale with angle
  force_vector[0].fy = restoring_force () * (force_vector[0].dy / h);

//cycle throguh all apertures looking for overlap with cone
  //printf ("cone radius = %f\n", r_cone);
  j = 1;
  for (i = 0; i <= ap; i++)
    {
//printf("%f\t%f\t%f\n",aperture[i].x, aperture[i].y,get_distance(next[0].x, next[0].y, aperture[i].x, aperture[i].y));
      if (get_distance (next[0].x, next[0].y, aperture[i].x, aperture[i].y) <
          r_cone)
        {
          h =
            get_distance (next[0].x, next[0].y, aperture[i].x, aperture[i].y);
//make sure accelerations are in the right direction
          force_vector[j].dx = aperture[i].x - next[0].x;
          force_vector[j].dy = aperture[i].y - next[0].y;
          force_vector[j].fx = get_air_force (aperture[i].hole_size) * (force_vector[j].dx /
h);    //scale with angle
          force_vector[j].fy =
            get_air_force (aperture[i].hole_size) * (force_vector[j].dy / h);

          printf
            (">>>Hit at aperture = (%f,%f), position = (%f,%f), force-x = %f  force-y = %f\
n",
             aperture[i].x, aperture[i].y, next[0].x, next[0].y, force_vector[j].fx,
             force_vector[j].fy);
          //printf ("\n>>>>>>>>dx = %f  dy = %f\n", force_vector[j].dx, force_vector[j].dy);
          j++;

        }

    }

}

void
sum_force_vectors ()
{
//create vectors scaled by force, add vectors to get summed force and direction
//we just divide all the x and y components of the vectors by force
//sum them from head to tail, then get the force direction and magnitude
```

```c
  int m;
  double sum_fx = 0;
  double sum_fy = 0;

  for (m = 0; m < 5; m++)
    {
      sum_fx = sum_fx + force_vector[m].fx;
      sum_fy = sum_fy + force_vector[m].fy;
    }
//get accelerations from force and mass, a = F/m
//and add to previous acceleration

  last_a[0].x = last_a[0].x + (sum_fx / pendulum_mass);
  last_a[0].y = last_a[0].y + (sum_fy / pendulum_mass);

  //printf ("sum_fx = %f    sum_fy = %f    sum_ax = %f    sum_ay = %f\n", sum_fx, sum_fy, last_a[0].x, last_a[0].y);

}

int
main (int argc, char *argv[])
{

  strcpy (fn1, "output.txt");
  printf ("\nOutput datafile selected = %s\n", fn1);

  if ((outdata = fopen (fn1, "w")) == NULL)
    {
      printf ("\nDatafile cannot be opened.\n");
      exit (1);
    }

  srand (time (NULL));

//set initial pendulum position and acceleration, number of apertures
  theta = random_angle ();
  initial_x = (aperture_array_radius / 2) * cos (theta);
  initial_y = (aperture_array_radius / 2) * sin (theta);
  next[0].x = initial_x;
  next[0].y = initial_y;
  last_a[0].x = 0;
  last_a[0].y = 0;
  num_apertures = apertures;

//check command line for initial x, initial y, and aperture file
  if (argc > 3)
    {
      initial_x = atof (argv[1]);
      initial_y = atof (argv[2]);
      next[0].x = initial_x;
      next[0].y = initial_y;

      strcpy (fn2, argv[3]);
      printf ("\nAperture file selected = %s\n", fn2);

      if ((indata = fopen (fn2, "r")) == NULL)
       {
         printf ("\nInput file cannot be opened.\n");
         exit (1);
       }

      fctr = 0;
      while (fgets (str, 64, indata) != NULL)
       {
//printf("%s",str);
         sscanf (str, "%lf\t%lf\t%lf", &x_dat, &y_dat, &h_dat);

         printf ("%f\t%f\t%f\n", x_dat, y_dat, h_dat);
```

```
              aperture[fctr].x = x_dat;
              aperture[fctr].y = y_dat;
              aperture[fctr].hole_size = h_dat;
              fctr++;
            }
          num_apertures = fctr - 1;
        }

//write parameters to file
    fprintf (outdata, "Parameter\tValue\tUnits\n");
    fprintf (outdata, "number of apertures = \t%d\n", num_apertures);
    fprintf (outdata, "pendulum length = \t%.7f\t%s\n", pendulum_length,
             "meters");
    fprintf (outdata, "radius of aperture array = \t%.7f\t%s\n",
             aperture_array_radius, "meters");
    fprintf (outdata, "pendulum mass = \t%.7f\t%s\n", pendulum_mass,
             "kilograms");
    fprintf (outdata, "cone radius = \t%.7f\t%s\n", r_cone, "meters");
    fprintf (outdata, "cone half angle = \t%d\t%s\n", cone_half_angle,
             "degrees");
    fprintf (outdata, "gravitational acceleration = \t%.7f\t%s\n", g,
             "meters/sec/sec");
    fprintf (outdata, "pendulum pressure = \t%d\t%s\n", plenum_pressure,
             "pascal");
    fprintf (outdata, "atmospheric pressure = \t%d\t%s\n", atomos_pressure,
             "pascal");
    fprintf (outdata, "air density = \t%.7f\t%s\n", air_density, "kg/m^3");
    fprintf (outdata, "delta-t = \t%.7f\t%s\n", tick, "seconds");
    fprintf (outdata, "pendulum initial x = \t%.7f\t%s\n", initial_x, "meters");
    fprintf (outdata, "pendulum initial y = \t%.7f\t%s\n", initial_y, "meters");
    fprintf (outdata, "\n\n");      //two lines will index GNUPlot data

    ap = 0;
    i = 0;

    if (argc < 3)
      {
        while (ap < num_apertures)
          {
            theta = random_angle ();
            radius = random_radius ();
//radius squared to ensure uniform density across area
            x_trial = radius * radius * cos (theta);
            y_trial = radius * radius * sin (theta);

//check_spacing returns zero if spacing is OK
            if (check_spacing (min_spacing, x_trial, y_trial, ap) == 0)
              {
                aperture[ap].x = x_trial;
                aperture[ap].y = y_trial;
                aperture[ap].hole_size = random_hole ();
                ap++;
                i++;
                if (i > attempts)
                  {
                    printf ("\n\n%s",
                            "***Minimum spacings appears to be too large ");
                    printf ("%s\n\n", "for number of apertures***");
                    break;
                  }
              }

          }

      }

//rescale aperture array to physical distances and save to file
//fprintf(outdata,"#x\ty\thole (m)\n");
    if (argc < 3)
      {
        for (ap = 0; ap < num_apertures; ap++)
          {
```

```c
            aperture[ap].x = aperture_array_radius * aperture[ap].x;
            aperture[ap].y = aperture_array_radius * aperture[ap].y;
            if (validation == TRUE)
               {
                 aperture[ap].hole_size = 0;       //no air flow for pendulum motion validation
               }
          }
      }

   for (ap = 0; ap < num_apertures; ap++)
      {
        fprintf (outdata, "%.7f\t%.7f\t%.7f\n", aperture[ap].x, aperture[ap].y,
              aperture[ap].hole_size);
      }

//print blank lines to file
   fprintf (outdata, "\n\n");      //two lines will index GNUPlot data

   printf ("\n%f\t%f\t%f\t%f\t%f\t%f\n", force_vector[0].fx,
         force_vector[0].fy, force_vector[0].dx, force_vector[0].dy,
         next[0].x, next[0].y);

   printf ("\nRestoring force = %f\n", restoring_force ());

   for (k = 0; k < trials; k++)
       {
//find and get all forces acting on the pendulum
        get_forces ();

//create vectors scaled by force, add vectors to get summed force and acceleration
        sum_force_vectors ();

//find new position after time tick
        next[0].x = next[0].x + (0.5 * last_a[0].x * tick * tick);
        next[0].y = next[0].y + (0.5 * last_a[0].y * tick * tick);

//print new (x,y) position to file and vector magnitude
        fprintf (outdata, "%.5f\t%.5f\t%.5f\n", next[0].x, next[0].y,
              sqrt ((next[0].x * next[0].x) + (next[0].y * next[0].y)));

       }

   fclose (outdata);

   time (&t);
   tmp = localtime (&t);
//format time using strftime, making certain we don't overwrite files
   while (strcmp (last_file_timestamp, file_timestamp) == 0)
      {
        strftime (file_timestamp, sizeof (file_timestamp), "%Y-%m-%d_%H-%M-%S",
              tmp);
      }

   strcpy (last_file_timestamp, file_timestamp);

   printf ("\nTime = %s\n", file_timestamp);

//make subfolder 'data' is it does not exist
   strcpy (sys_cmd, "mkdir -p data");
   retval = system (sys_cmd);
   if (retval == 0)
      {
        printf ("Folder 'data' exists\n");
      }
   if (retval != 0)
      {
        printf ("Folder 'data' created\n");
      }

   if (plot == TRUE)
      {
//create plot using gnuplot
```

```c
      retval = system ("gnuplot graph.gnu");

//make an archive file of graph
      strcpy (sys_cmd, "cp ");
      strcat (sys_cmd, "graph.png data/graph_");
      strcat (sys_cmd, file_timestamp);
      strcat (sys_cmd, ".png");
      retval = system (sys_cmd);
    }

//make an archive file of the datafile
  strcpy (sys_cmd, "cp ");
  strcat (sys_cmd, "output.txt data/data_");
  strcat (sys_cmd, file_timestamp);
  strcat (sys_cmd, ".txt");
  retval = system (sys_cmd);

  printf ("Done\n");
  return (0);
}
```

## Appendix IV – GNUPlot Script (graph.gnu)

```
set terminal pngcairo  enhanced font "arial,16" fontscale 1.0 size 750, 750
set output 'graph.png'
set grid nopolar
set grid xtics nomxtics noytics nomytics noztics nomztics \
 nox2tics nomx2tics y2tics nomy2tics nocbtics nomcbtics
set grid layerdefault   linetype -1 linecolor rgb "gray"  linewidth 0.200
unset key
set autoscale x
set autoscale y
set title "Pneumatic Chaotic Pendulum"
set xlabel "X"
set ylabel "Y"
plot 'output.txt' skip 17 index 1 u 1:2 w lines lt rgb "blue", 'output.txt' skip 17 index 0 u 1:2 pt 7 lc rgb "red" ps 1
```

## Appendix V – Initial Conditions

| Number of Apertures | 20 | |
|---|---|---|
| Pendulum Length | 0.445 | meters |
| Aperture Array Radius | 0.25 | meters |
| Pendulum Mass | 0.009 | kilograms |
| Cone Radius | 0.019 | meters |
| Cone Apex Half Angle | 30 | degrees |
| Gravitation Acceleration (g) | 9.8 | meters/sec/sec |
| Plenum Pressure | 102500 | pa |
| Atmospheric Pressure | 101000 | pa |
| Air Density | 1.225 | kg/m^3 |